\documentclass[letterpaper, 10 pt, conference]{ieeeconf}
\IEEEoverridecommandlockouts
\usepackage{cite}
\usepackage{amsmath}
\usepackage{amsfonts}
\usepackage{graphicx}
\usepackage{url}
\let\labelindent\relax
\usepackage{enumitem}
\usepackage{microtype}

\newtheorem{theorem}{Theorem}
\newtheorem{remark}{Remark}

\usepackage{xcolor}

\usepackage{tikz}
\usetikzlibrary{arrows.meta, positioning, fit, calc}

\title{\LARGE \bf NDKF: A Neural-Enhanced Distributed Kalman Filter \\ for Nonlinear Multi-Sensor Estimation}
\author{Siavash Farzan and Bennett Parisi%
\thanks{The authors are with the Electrical Engineering Department, California Polytechnic State University, San Luis Obispo, CA 93407, USA.}
\thanks{Corresponding author: {\tt\footnotesize sfarzan@calpoly.edu}}
}

\begin{document}

\maketitle
\thispagestyle{empty}
\pagestyle{empty}

\begin{abstract}
We propose a Neural-Enhanced Distributed Kalman Filter (NDKF) for multi-sensor state estimation in nonlinear systems. Unlike traditional Kalman filters that rely on explicit analytical models and assume centralized fusion, NDKF leverages neural networks to replace analytical process and measurement models with learned mappings while each node performs local prediction and update steps and exchanges only compact posterior summaries with its neighbors. This distributed design reduces communication overhead and avoids a central fusion bottleneck. We provide sufficient mean-square stability conditions under bounded Jacobians and well-conditioned innovations, together with practically checkable proxies such as Jacobian norm control and innovation monitoring. We also discuss consistency under learned-model mismatch, including covariance inflation and covariance-intersection fusion when cross-correlations are uncertain.
Simulations on a 2D nonlinear system with four partially observing nodes show that NDKF
outperforms a distributed EKF baseline under model mismatch and yields improved estimation
accuracy with modest communication requirements.
\end{abstract}

\section{Introduction}
\label{sec:introduction}

State estimation lies at the heart of a wide range of applications in robotics, sensor networks, and control systems. Classical Kalman filtering techniques remain the gold standard for linear-Gaussian problems; however, many real-world scenarios present nonlinear dynamics or complex measurement functions that challenge traditional approaches. Moreover, when multiple agents operate in a distributed fashion, exchanging all raw data can impose high communication overhead and hinder real-time operation.

Distributed Kalman filtering has long been explored to enable multi-sensor state estimation without relying on a centralized fusion node. Early work on consensus-based filters \cite{Olfati2005} and \cite{Olfati2007}, and on federated approaches \cite{Carlson1994}, allowed individual nodes to update local estimates and share compact summaries (such as means and covariances) with their neighbors. While effective for linear or mildly nonlinear systems, these methods require careful coordination to avoid overconfidence and inconsistent estimates \cite{Carli2007}.

To handle nonlinearities and model uncertainties, extensions of the standard Kalman filter have been developed. Variants of the Extended Kalman Filter (EKF) \cite{Jazwinski2007} approximate local Jacobians to manage nonlinear dynamics, with established stochastic stability conditions under boundedness assumptions \cite{Reif1999}. Unscented Kalman filters (UKF) \cite{Julier2004} employ deterministic sampling to capture higher-order moments without linearization. Both approaches have been adapted for distributed settings \cite{Battistelli2016} and \cite{Li2016}, though they depend on explicit system models that may not fully capture the complexities of real-world dynamics.
Robust fusion techniques such as covariance intersection \cite{Julier2017,Hu2012} address unknown cross-correlations and guard against overconfident fusion; they are complementary to, rather than substitutes for, local outlier rejection mechanisms.

In parallel, the integration of machine learning into filtering frameworks has led to neural-based Kalman filters. Examples include deep latent state-space models \cite{Krishnan2015} and \cite{Fraccaro2016}, and KalmanNet-style gain learning \cite{Revach2022}. The broader lineage also includes dual Kalman filtering ideas \cite{NIPS1996DualKF} and recent multiagent nonlinear filtering-and-learning analyses \cite{MLSP2024NonlinearFiltering}. However, most neural filtering approaches are applied in centralized settings or assume linear models at individual nodes. Extending them to distributed, partially observed multi-agent systems raises concrete challenges: ensuring stability despite learned-model bias, preventing overconfidence under unknown cross-correlations, and managing the cost of repeated neural evaluations and Jacobians in real time.

NDKF targets these gaps. (i) Each node learns its \emph{own} measurement mapping while the process mapping can be learned from shared logs or simulation; nodes exchange only posterior summaries. (ii) Information-form consensus provides a principled fusion route and admits a robust covariance-intersection fallback when cross-correlations are unknown. (iii) Our analysis supplies sufficient stability conditions under local linearization with Lipschitz-bounded neural Jacobians and introduces \emph{practically checkable proxies} (e.g., innovation conditioning and Jacobian norm control) that can be verified or enforced during deployment. We do not assume parameter sharing; here $\theta$ is a collective placeholder for all trainable parameters (process and per-node measurement), and only posterior summaries (not $\theta$) are communicated.

Our work bridges these research streams by proposing a Neural-Enhanced Distributed Kalman Filter (NDKF). NDKF integrates neural network approximations of both system dynamics and measurement functions with a consensus-based distributed fusion mechanism. Unlike centralized methods such as KalmanNet, NDKF enables each sensor node to process local measurements and share only concise state summaries (e.g., local estimates and covariances), which reduces communication overhead and preserves data privacy. This framework addresses scalability, model uncertainty, and nonlinearities in a fully distributed manner where each node observes a partial, potentially nonlinear function of the system state, and provides a robust alternative to conventional EKF/UKF-based methods in multi-sensor applications.

The primary contributions of this work are:
\begin{itemize}[leftmargin=12pt]
    \item[i.] An integrated filtering framework combining neural network approximations of system dynamics and measurement functions with distributed Kalman filtering to effectively address complex nonlinear behaviors.
    \item[ii.] A consensus-based distributed fusion mechanism to enable scalable, decentralized state estimation, complemented by a detailed computational complexity analysis and an optional covariance-intersection fallback when cross-correlations are unknown.
    \item[iii.] A mean-square stability and convergence analysis for the NDKF under local linearization and Lipschitz continuity, together with practically checkable proxies for deployment.
    \item[iv.] Validation through simulation studies on a 2D multi-sensor scenario, demonstrating superior performance of the proposed NDKF compared to a conventional distributed EKF baseline.
\end{itemize}

\subsection*{Practical Training Without Ground-Truth States}
A common concern is that learning $f_{\theta}$ and $h_{\theta,i}$ requires state labels that are unavailable at deployment. In NDKF, we use regimes that avoid this requirement:
\begin{itemize}[leftmargin=*]
    \item \textit{Supervised via proxies:} brief calibration runs with external instrumentation or high-fidelity simulators (``digital twins'') provide approximate state labels to pretrain $f_{\theta}$ and $h_{\theta,i}$.
    \item \textit{Latent/self-supervised:} treat $\mathbf{x}_k$ as latent and train by maximizing predictive likelihood or minimizing multi-sensor innovation residuals over sequences, leveraging \cite{Krishnan2015}, \cite{Fraccaro2016}, and \cite{Revach2022} as building blocks. No state labels are required.
    \item \textit{Residual learning:} start from a nominal model $f_0,h_{0,i}$ and learn small corrections $\Delta f, \Delta h_i$ using pseudo-labels from high-trust smoothers or consensus posteriors; regularize the residuals to preserve stability.
\end{itemize}
These regimes keep training feasible when states are hidden, and align with the distributed nature of our method: nodes train $h_{\theta,i}$ locally, while $f_{\theta}$ can be learned centrally or from simulated traces and then deployed.

\subsection*{Representative Deployment Scenarios}
NDKF is particularly appealing in settings where sensors are spatially distributed, measurements are heterogeneous, and accurate first-principles models are unavailable.
Examples include cooperative ground or aerial robots with complementary sensing, environmental monitoring with mixed acoustic/range/chemical sensors, intelligent transportation systems with roadside and onboard measurements, and industrial process monitoring across multiple stations. In such deployments, the main practical challenges are communication delays or packet drops, time synchronization, sensor bias drift, occasional outliers or faults, and limited on-node compute. The present design addresses the communication burden by exchanging posterior summaries rather than raw data, and it can be combined with innovation gating and covariance inflation to improve resilience under field conditions~\cite{Cline2026}.

\section{NDKF Framework}
\label{sec:methodology}

In this section, we describe the proposed Neural-Enhanced Distributed Kalman Filter (NDKF) framework in detail.

\subsection{System Setup}
\label{sec:system-setup}

We consider a dynamical system with state vector $\mathbf{x}_k{\in}\mathbb{R}^n$ at discrete time $k$. The evolution of the state is modeled by:
\begin{equation}
    \mathbf{x}_{k+1} = f_{\theta_f}\bigl(\mathbf{x}_k,\mathbf{u}_k\bigr) + \mathbf{w}_k,
    \label{eq:system-dynamics}
\end{equation}
where $f_{\theta_f}(\cdot,\cdot)$ is a potentially nonlinear function approximated by a
neural network with parameters $\theta_f$, $\mathbf{u}_k$ denotes an optional control input,
and $\mathbf{w}_k$ is zero-mean process noise with covariance matrix
$\mathbf{Q}\succeq \mathbf{0}$. In the benchmark considered in Section~\ref{sec:results},
the system is autonomous and we therefore set $\mathbf{u}_k=\mathbf{0}$ for simplicity.

We employ a network of $N$ sensor nodes, each collecting noisy measurements of the state. Node $i \in \{1,2,\dots,N\}$ obtains
\begin{equation}
    \mathbf{y}_{k,i} = h_{\theta_{h,i},i}\bigl(\mathbf{x}_k\bigr) + \mathbf{v}_{k,i},
    \label{eq:measurement-model}
\end{equation}
where $h_{\theta_{h,i},i}(\cdot)$ is the local measurement function for node $i$, represented
or assisted by a neural network with parameters $\theta_{h,i}$, and $\mathbf{v}_{k,i}$ is
zero-mean measurement noise with covariance $\mathbf{R}_i\succeq \mathbf{0}$.
Depending on the application, each $h_{\theta,i}(\cdot)$ may observe a subset of the state or a particular nonlinear transformation of $\mathbf{x}_k$.
Note that the parameter sets $\theta$ used in $f_{\theta}(\cdot)$ and $h_{\theta,i}(\cdot)$ are not necessarily the same and can be independently trained neural networks with distinct parameters.

Here, the process noise $\mathbf{w}_k$ and the measurement noise $\mathbf{v}_{k,i}$ are assumed to be zero-mean and temporally white (i.e., uncorrelated across time steps). Moreover, these noise processes are mutually independent, with covariances $\mathbf{Q}$ and $\mathbf{R}_i$, respectively.

To enable efficient estimation, we assume nodes can exchange compact information (e.g., local estimates and covariance matrices) over a communication network, but we do not require that all raw measurements be sent to a central fusion center. The goal is to estimate $\mathbf{x}_k$ at each time step while preserving distributed operation and efficiently handling nonlinear dynamics through neural network approximations.

\begin{remark} \textit{(Parameterization, control inputs, and communication).}
We use $\theta_f$ for the dynamics network and $\theta_{h,i}$ for the measurement network
at node $i$; no parameter sharing is assumed unless explicitly stated. When a nominal affine
control channel is known, one may also write
\[
\mathbf{x}_{k+1}=f_{\theta_f}(\mathbf{x}_k)+\mathbf{B}_k\mathbf{u}_k+\mathbf{w}_k.
\]
During filtering, nodes exchange only posterior summaries (means/covariances or information
pairs), not raw measurements and not neural parameters.
\end{remark}

\subsection{Neural Network for System and Measurement Modeling}
\label{sec:nn-modeling}

To handle the potentially nonlinear nature of the system dynamics \eqref{eq:system-dynamics} and measurement functions \eqref{eq:measurement-model}, we employ neural networks to approximate $f_{\theta}(\cdot)$ and $h_{\theta,i}(\cdot)$. These networks are parameterized by $\theta$, which can be learned from historical data or adapted online.

We represent the mapping $\mathbf{x}_k \mapsto \mathbf{x}_{k+1}$ through a neural network $f_{\theta}(\cdot)$. A typical architecture might include multiple fully connected layers with nonlinear activation functions, although any suitable architecture (e.g., convolutional or recurrent layers) can be used depending on domain requirements. The network parameters $\theta$ are determined by minimizing a loss function
\begin{equation}
    \mathcal{L}_{\mathrm{dyn}}\bigl(\theta\bigr) \;=\; \sum_{j=1}^{M}\,
    \bigl\|\mathbf{x}_{k+1}^{(j)} \;-\; f_{\theta}\bigl(\mathbf{x}_k^{(j)}\bigr)\bigr\|^2,
    \label{eq:dyn-loss}
\end{equation}
Here, the training samples $\{(\mathbf{x}_k^{(j)}, \mathbf{x}_{k+1}^{(j)})\}_{j=1}^{M}$ are obtained from historical measurements or high-fidelity simulations.
The minimization of \eqref{eq:dyn-loss} ensures the network accurately models the underlying system transitions.
In scenarios where the true system dynamics are unknown, one may employ an approximate model or direct sensor observations during an initial phase to generate these samples and to enable the neural network to learn a corrective (residual) representation of the dynamics.

Each sensor node $i$ can optionally employ a neural network $h_{\theta,i}(\cdot)$ to capture the (possibly nonlinear) relationship between the state $\mathbf{x}_k$ and the measurements $\mathbf{y}_{k,i}$. Similar to the dynamics network, $h_{\theta,i}(\cdot)$ can be trained by minimizing
\begin{equation}
    \mathcal{L}_{\mathrm{meas},i}\bigl(\theta\bigr) \;=\; \sum_{j=1}^{M_i}\,
    \bigl\|\mathbf{y}_{k,i}^{(j)} \;-\; h_{\theta,i}\bigl(\mathbf{x}_k^{(j)}\bigr)\bigr\|^2,
    \label{eq:meas-loss}
\end{equation}
where $(\mathbf{x}_k^{(j)}, \mathbf{y}_{k,i}^{(j)})$ pairs are the measurement data specific to node $i$. When the measurement function is linear or partially known, $h_{\theta,i}(\cdot)$ can be reduced to a simpler parametric form or omitted in favor of a standard linear observation model.

In many applications, model training occurs \emph{offline} using a representative dataset, after which the parameters $\theta$ are fixed for the subsequent filtering process. Alternatively, in cases where the system evolves over time or new operating regimes appear, an \emph{online} training approach can be adopted. This may involve periodically updating $\theta$ using fresh data to refine the learned models $f_{\theta}$ and $h_{\theta,i}$ and maintain accuracy under changing conditions.

To mitigate distribution shift, we (i) learn residual corrections on top of nominal models when available, (ii) regularize networks for smoothness (spectral/weight decay) to control Jacobians, (iii) use early stopping and multiple random initializations to avoid poor local minima, and (iv) enable lightweight online adaptation of small residual heads if the operating regime drifts, while keeping the core model fixed to preserve stability.

\begin{figure}[!tb]
    \centering
    \vskip 5pt
    \begin{tikzpicture}[
    font=\scriptsize,
    >=Latex,
    node distance=4mm and 5mm,
    block/.style={
        draw,
        rounded corners,
        thick,
        align=center,
        fill=gray!10,
        minimum height=8mm,
        text width=3.05cm,
        inner sep=4pt
    },
    model/.style={
        draw,
        rounded corners,
        thick,
        align=center,
        fill=blue!8,
        minimum height=8mm,
        text width=2.65cm,
        inner sep=4pt
    },
    io/.style={
        draw,
        rounded corners,
        align=center,
        fill=green!8,
        minimum height=8mm,
        text width=1.45cm,
        inner sep=3pt
    },
    msg/.style={
        draw,
        rounded corners,
        align=center,
        fill=orange!10,
        minimum height=8mm,
        text width=1.55cm,
        inner sep=3pt
    },
    flow/.style={->, thick}
]

\node[block] (pred) {Prediction\\
use $f_{\theta_f}$ and $\mathbf{F}_{\theta_f,k}$\\
compute $\hat{\mathbf{x}}_{k+1|k},\ \mathbf{P}_{k+1|k}$};

\node[block, below=of pred] (upd) {Local update at node $i$\\
use $h_{\theta_{h,i},i}$ and $\mathbf{H}_{\theta_{h,i},i,k}$\\
compute $\hat{\mathbf{x}}_{k+1|k+1,i},\ \mathbf{P}_{k+1|k+1,i}$};

\node[block, below=of upd] (pack) {Information packaging\\
$\mathbf{W}_i=\mathbf{P}_{k+1|k+1,i}^{-1}$\\
$\mathbf{z}_i=\mathbf{W}_i\hat{\mathbf{x}}_{k+1|k+1,i}$};

\node[block, below=of pack] (fusion) {Neighbor fusion\\
exchange $(\mathbf{W}_j,\mathbf{z}_j)$\\
consensus or covariance intersection};

\node[io, below=of fusion, text width=2.7cm] (post) {Output fused posterior\\
$\hat{\mathbf{x}}_{k+1|k+1,i}^{(\mathrm{fusion})}$\\
$\mathbf{P}_{k+1|k+1,i}^{(\mathrm{fusion})}$};

\node[io, left=of pred] (prior) {Input posterior\\
$\hat{\mathbf{x}}_{k|k,i}$\\
$\mathbf{P}_{k|k,i}$};

\node[model, above=of pred] (dyn) {Shared dynamics NN\\
$f_{\theta_f}(\mathbf{x},\mathbf{u})$\\
{\tiny 3 hidden layers, 128 units}};

\node[model, right=of upd, text width=1.75cm,yshift=0.5cm] (measnn) {Node-$i$\\
meas. NN\\
$h_{\theta_{h,i},i}$};

\node[io, below=0.3mm of measnn, text width=1.75cm] (measy) {Local\\
measurement\\
$\mathbf{y}_{k+1,i}$};

\node[msg, right=of fusion, text width=2.05cm] (msgs) {Neighbor messages\\
$\{(\mathbf{W}_j,\mathbf{z}_j)\}_{j\in\mathcal{N}_i}$};

\draw[flow] (msgs) to (fusion);

\node[
    draw,
    dashed,
    rounded corners,
    inner sep=4pt,
    fit=(pred)(upd)(pack)(fusion),
    label={[font=\tiny,xshift=1.1cm,yshift=-0.1cm]Node $i$ runtime pipeline}
] (runtime) {};

\draw[flow] (prior) to (pred);
\draw[flow] (dyn) to (pred);
\draw[flow] (pred) to (upd);
\draw[flow] (upd) to (pack);
\draw[flow] (pack) to (fusion);
\draw[flow] (fusion) to (post);

\draw[flow] (measnn.west) to (upd.east);
\draw[flow] (measy.west) to (upd.east);

\node[align=center, font=\tiny] at ($(fusion.south)+(0,-2.5mm)$)
{Only posterior summaries are \;\; exchanged across neighbors\;\;};

\node[align=center, font=\tiny] at ($(post.west)+(-1.3cm,-0.25cm)$)
{used as input posterior\\for the next step};

\node [coordinate, left of=prior, node distance=1.2cm] (coord1) {};
\draw[dashed] (post.west) -| (coord1);
\draw[->,dashed] (coord1) -- (prior.west);

\end{tikzpicture}
    \caption{NDKF overview. Each node executes prediction using the shared dynamics network $f_{\theta_f}$, applies a local update with its own measurement network $h_{\theta_{h,i},i}$, packages the posterior into information form $(\mathbf{W}_i,\mathbf{z}_i)$, and fuses neighbor messages through consensus or covariance intersection. Raw measurements remain local.}
    \label{fig:ndkf-architecture}
\end{figure}

\subsection{Kalman Filter with Neural Network Integration}
\label{sec:kf-steps}

The proposed filter preserves the classical two-stage prediction/update but \emph{differs} in three practical aspects: (i) the state transition and per-node observation functions are \emph{learned} (and may be nonlinear and heterogeneous across nodes); (ii) Jacobians $\mathbf{F}_{\theta,k}$ and $\mathbf{H}_{\theta,i,k}$ come from autodiff of these networks or finite differences; and (iii) the local posteriors are fused in information form (with an optional CI fallback) to mitigate overconfidence when cross-correlations are unknown.

\subsubsection{Prediction Step}
Let $\hat{\mathbf{x}}_{k|k}$ and $\mathbf{P}_{k|k}$ denote the estimated mean and covariance of the state at time $k$ (after processing the measurements up to $k$). The prediction step for time $k{+}1$ uses the learned dynamics function $f_{\theta}$ from \eqref{eq:system-dynamics}:\looseness=-1
\begin{equation}
    \hat{\mathbf{x}}_{k+1|k} \;=\; f_{\theta}\bigl(\hat{\mathbf{x}}_{k|k}\bigr),
    \label{eq:kf-prediction-state}
\end{equation}
\begin{equation}
    \mathbf{P}_{k+1|k} \;=\; \mathbf{F}_{\theta,k}\,\mathbf{P}_{k|k}\,\mathbf{F}_{\theta,k}^{T} \;+\; \mathbf{Q},
    \label{eq:kf-prediction-cov}
\end{equation}
where $\mathbf{F}_{\theta,k}$ is the Jacobian of $f_{\theta}(\cdot)$ evaluated at $\hat{\mathbf{x}}_{k|k}$. In practice, $\mathbf{F}_{\theta,k}$ can be obtained via automatic differentiation or finite differences. The matrix $\mathbf{Q}$ is the process noise covariance, which may be tuned empirically or even learned if sufficient data and model structures are available.

\subsubsection{Local Measurement Update}
Each node $i$ receives a measurement $\mathbf{y}_{k,i}$ according to \eqref{eq:measurement-model}, which relies on the learned observation function $h_{\theta,i}(\cdot)$. Defining $\mathbf{H}_{\theta,i,k}$ as the Jacobian of $h_{\theta,i}(\cdot)$ at the predicted state, the standard Kalman innovation form applies:
\begin{equation}
    \mathbf{S}_{k+1,i} \;=\; \mathbf{H}_{\theta,i,k+1}\,\mathbf{P}_{k+1|k}\,\mathbf{H}_{\theta,i,k+1}^{T} \;+\; \mathbf{R}_i,
    \label{eq:kf-innovation-cov}
\end{equation}
\begin{equation}
    \mathbf{K}_{k+1,i} \;=\; \mathbf{P}_{k+1|k}\,\mathbf{H}_{\theta,i,k+1}^{T}\,\mathbf{S}_{k+1,i}^{-1},
    \label{eq:kf-kalman-gain}
\end{equation}
\begin{equation}
    \hat{\mathbf{x}}_{k+1|k+1,i} = \hat{\mathbf{x}}_{k+1|k} + \mathbf{K}_{k+1,i}\,\Bigl[\mathbf{y}_{k+1,i} - h_{\theta,i}\bigl(\hat{\mathbf{x}}_{k+1|k}\bigr)\Bigr],
    \label{eq:kf-meas-update-state}
\end{equation}
\begin{equation}
    \mathbf{P}_{k+1|k+1,i} \;=\; \bigl(\mathbf{I} - \mathbf{K}_{k+1,i}\,\mathbf{H}_{\theta,i,k+1}\bigr)\,\mathbf{P}_{k+1|k}.
    \label{eq:kf-meas-update-cov}
\end{equation}

In \eqref{eq:kf-innovation-cov}--\eqref{eq:kf-meas-update-cov}, the subscript $(k{+}1|k)$ indicates the predicted quantities from \eqref{eq:kf-prediction-state}--\eqref{eq:kf-prediction-cov}, and $\mathbf{R}_i$ is the measurement noise covariance at node $i$. If $h_{\theta,i}(\cdot)$ is linear or partially known, $\mathbf{H}_{\theta,i,k+1}$ can be simplified accordingly.

\paragraph*{Relation to distributed EKF}
The algebraic prediction/update equations in NDKF intentionally retain the EKF structure; the novelty lies in \emph{what} is linearized and \emph{how} local posteriors are fused. Specifically, (i) the process and measurement maps are learned from data rather than assumed analytically
known, (ii) each node may carry a distinct learned measurement model, and (iii) fusion is performed in information form with a consistency-preserving covariance-intersection fallback when cross-correlations are uncertain. Consequently, if the learned models are replaced by exact analytical ones and covariance intersection is not needed, NDKF reduces locally to a distributed EKF/information filter. We emphasize that NDKF does not eliminate linearization error; rather, it addresses a different failure mode, namely inaccurate analytical process and measurement models. When linearization error dominates, derivative-free distributed filters may
be preferable.

\subsection{Distributed Fusion Mechanism}
\label{sec:distributed-fusion}

Once each node $i$ completes its local measurement update and obtains $\hat{\mathbf{x}}_{k|k,i}$ and $\mathbf{P}_{k|k,i}$, these partial estimates must be combined to form a globally consistent state estimate. Unlike standard distributed Kalman filters that assume explicit process and measurement models, our neural-based approach still relies on exchanging only concise summary information (e.g., posterior mean and covariance) to achieve consensus without centralizing raw data or neural parameters.

A straightforward method to merge local estimates is via their \emph{information form}. Each node $i$ transforms its posterior covariance $\mathbf{P}_{k|k,i}$ into the precision (inverse covariance) matrix, $\mathbf{P}_{k|k,i}^{-1}$, and computes the weighted contribution of its local state estimate $\hat{\mathbf{x}}_{k|k,i}$:
\begin{equation}
    \mathbf{W}_i \;=\; \mathbf{P}_{k|k,i}^{-1}, 
    \qquad
    \mathbf{z}_i \;=\; \mathbf{W}_i\,\hat{\mathbf{x}}_{k|k,i}.
    \label{eq:info-fusion-contrib}
\end{equation}
Each node then aggregates the information from its neighbors (or potentially all other nodes, depending on the network topology). Denoting the set of nodes that communicate with node $i$ at time $k$ by $\mathcal{N}_i$, the local fusion update may be written as:
\begin{equation}
    \mathbf{W}_{k|k,i}^{(\mathrm{fusion})} \;=\; \sum_{j \in \mathcal{N}_i} \mathbf{W}_j,
    \qquad
    \mathbf{z}_{k|k,i}^{(\mathrm{fusion})} \;=\; \sum_{j \in \mathcal{N}_i} \mathbf{z}_j.
    \label{eq:info-fusion-aggregation}
\end{equation}
The fused covariance is then:
\begin{equation}
    \mathbf{P}_{k|k,i}^{(\mathrm{fusion})} \;=\; \bigl(\mathbf{W}_{k|k,i}^{(\mathrm{fusion})}\bigr)^{-1},
    \label{eq:fusion-P}
\end{equation}
and the fused state estimate at node $i$ becomes:
\begin{equation}
    \hat{\mathbf{x}}_{k|k,i}^{(\mathrm{fusion})} \;=\; 
    \mathbf{P}_{k|k,i}^{(\mathrm{fusion})}\,\mathbf{z}_{k|k,i}^{(\mathrm{fusion})}.
    \label{eq:fusion-x}
\end{equation}
Although \eqref{eq:info-fusion-aggregation}--\eqref{eq:fusion-x} represent a single fusion step, repeated local averaging can be performed if network communication is limited or if nodes reside in different subgraphs. Such \emph{consensus} approaches allow the global estimate to converge asymptotically to a fully fused solution, even without a central controller.

By exchanging only $\hat{\mathbf{x}}_{k|k,i}$ and $\mathbf{P}_{k|k,i}$, the NDKF avoids sending large amounts of sensor data, which is particularly beneficial given the learned neural models at each node. This lowers communication overhead and protects measurement privacy. Additionally, local computation scales more favorably than a central architecture, since each node only processes its own and its neighbors’ estimates.

When cross-correlations between local posteriors are unknown, \eqref{eq:info-fusion-aggregation} can be replaced by \emph{covariance intersection} (CI) \cite{Julier2017, Prabhakar2025}: 
$\mathbf{W}_{\mathrm{CI}}=\omega \mathbf{W}_a + (1-\omega)\mathbf{W}_b$, 
$\mathbf{z}_{\mathrm{CI}}=\omega \mathbf{z}_a + (1-\omega)\mathbf{z}_b$, $\omega\in[0,1]$. 
CI guarantees consistency (non-overconfident fusion) at the cost of conservatism and integrates seamlessly with our information-form exchange.

\subsection{Algorithm 1: One NDKF step at node $i$}\label{subsec:alg}
Here we summarize the per-time-step procedure executed at node \(i\), covering prediction, local measurement update, information-form packaging, and consensus fusion (with an optional covariance-intersection fallback), mapping inputs \(\hat{\mathbf{x}}_{k|k}, \mathbf{P}_{k|k}, \mathbf{y}_{k+1,i}\) and neighbors' messages to the fused posterior \(\hat{\mathbf{x}}_{k+1|k+1,i}^{(\mathrm{fusion})}\) and \(\mathbf{P}_{k+1|k+1,i}^{(\mathrm{fusion})}\).
\begin{enumerate}[leftmargin=*]
\item \textit{Prediction:} $\hat{\mathbf{x}}_{k+1|k} \leftarrow f_{\theta}\!\left(\hat{\mathbf{x}}_{k|k}\right)$, 
$\mathbf{P}_{k+1|k} \leftarrow \mathbf{F}_{\theta,k}\mathbf{P}_{k|k}\mathbf{F}_{\theta,k}^{\!\top}+\mathbf{Q}$, 
with $\mathbf{F}_{\theta,k} := \nabla f_{\theta}(\hat{\mathbf{x}}_{k|k})$ (autodiff or finite differences).
\item \textit{Local update:} form $\mathbf{H}_{\theta,i,k+1} := \nabla h_{\theta,i}(\hat{\mathbf{x}}_{k+1|k})$, innovation covariance $\mathbf{S}_{k+1,i}$, gain $\mathbf{K}_{k+1,i}$ via \eqref{eq:kf-innovation-cov}--\eqref{eq:kf-kalman-gain}, and update \eqref{eq:kf-meas-update-state}--\eqref{eq:kf-meas-update-cov}.
\item \textit{Information packaging:} $\mathbf{W}_i \leftarrow \mathbf{P}_{k+1|k+1,i}^{-1}$, 
$\mathbf{z}_i \leftarrow \mathbf{W}_i \hat{\mathbf{x}}_{k+1|k+1,i}$.
\item \textit{Consensus fusion:} aggregate $\sum_{j \in \mathcal{N}_i}\mathbf{W}_j, \sum_{j \in \mathcal{N}_i}\mathbf{z}_j$ to obtain fused $\mathbf{P}_{k+1|k+1,i}^{(\mathrm{fusion})}$ and $\hat{\mathbf{x}}_{k+1|k+1,i}^{(\mathrm{fusion})}$ via \eqref{eq:fusion-P}--\eqref{eq:fusion-x}. \emph{Optional:} if cross-correlations are unknown, use covariance intersection in place of \eqref{eq:info-fusion-aggregation}.
\end{enumerate}

\subsection{Initialization and Parameter Tuning}
\label{sec:init-param-tuning}

A successful deployment of the proposed NDKF framework relies on careful initialization of the filter state and covariances, as well as proper selection of neural network and filter hyperparameters.
Each node $i$ requires an initial state estimate $\hat{\mathbf{x}}_{0|0,i}$ and covariance $\mathbf{P}_{0|0,i}$. In the simplest case where little prior information is available, $\hat{\mathbf{x}}_{0|0,i}$ can be set to a zero vector and $\mathbf{P}_{0|0,i}$ chosen as a large diagonal matrix to reflect high uncertainty. If partial domain knowledge exists (e.g., an approximate position or measurement from an external reference), the initialization can leverage that information for a more accurate starting point. For consistency across nodes, it may be beneficial to align all $\mathbf{P}_{0|0,i}$ to a common baseline, unless certain nodes are known to have superior initial estimates.

The architectures of $f_{\theta}(\cdot)$ and, when applicable, $h_{\theta,i}(\cdot)$ are defined by choices of layer width, depth, and activation functions. We fix these hyperparameters based on preliminary experiments that balance approximation accuracy and computational efficiency.

Process noise $\mathbf{Q}$ and measurement noise $\mathbf{R}_i$ often require tuning to match real-world conditions. In practice, these matrices can be set by empirical measurement of sensor noise characteristics or process variability, or trial-and-error based on the observed filter performance (e.g., tracking accuracy), or even data-driven estimates, where additional parameters in $f_{\theta}$ or $h_{\theta,i}$ learn noise levels adaptively.

If automatic differentiation is used to compute Jacobians $\mathbf{F}_{\theta,k}$ and $\mathbf{H}_{\theta,i,k}$, one must ensure the chosen framework (e.g., TensorFlow, PyTorch) supports gradients for all operations involved in $f_{\theta}$ and $h_{\theta,i}$. When gradients are unavailable or prohibitively expensive to compute, finite-difference approximations offer an alternative, but at a higher computational cost.
Ensuring consistency in initialization and careful hyperparameter tuning enables the proposed filter to converge more rapidly, adapt to changing conditions, and maintain reliable estimates across multiple sensor nodes without centralizing measurements.

\subsection{Scalability and Computational Complexity}
\label{sec:computational-complexity}

A key consideration in deploying the proposed NDKF framework across multiple sensor nodes is the per-iteration computational load at each node. We outline below the dominant factors that contribute to the overall complexity.

At each prediction step, the function $f_{\theta}(\cdot)$ (or $h_{\theta,i}(\cdot)$, if used in the measurement function) must be evaluated. If the network has $L$ layers, with $n_\mathrm{in}$ inputs, $n_\mathrm{out}$ outputs, and an average of $n_\mathrm{h}$ neurons per hidden layer, the forward pass typically requires on the order of
\begin{equation}
    O\bigl(n_\mathrm{in} \, n_\mathrm{h} \;+\; (L-2) \, n_\mathrm{h}^2 \;+\; n_\mathrm{h} \, n_\mathrm{out}\bigr)
\end{equation}
basic arithmetic operations. For high-dimensional states or more complex neural architectures, this term may dominate the per-step computation.

To integrate the neural network outputs into the Kalman filter, the Jacobians $\mathbf{F}_{\theta,k}$ and $\mathbf{H}_{\theta,i,k}$ are needed for the covariance prediction and measurement update, respectively. Depending on implementation, they can be obtained via either \emph{automatic differentiation} or \emph{finite differences}.
Automatic differentiation is applicable if a framework such as PyTorch or TensorFlow is used, and the computational overhead is roughly proportional to a second pass through the network’s layers in order to maintain and backpropagate gradients.
On the other hand, finite differences requires multiple forward passes of the neural network for each state dimension, typically $2n$ evaluations for an $n$-dimensional state, rendering it $O(n)$ times more expensive than a single forward pass.
The choice between these methods thus influences per-step computational costs and memory usage.

For each node $i$, the standard Kalman filter update in \eqref{eq:kf-meas-update-state}--\eqref{eq:kf-meas-update-cov} involves matrix multiplications and inversions whose complexity depends on the state dimension $n$ and the measurement dimension $m_i$. Generally, matrix inversion or factorization can be $O(n^3)$ if $\mathbf{P}_{k|k-1}$ is $n\times n$, though efficient numerical methods (e.g., Cholesky decomposition~\cite{Krishnamoorthy2013}) can reduce constants in practice.

Each fusion round (as in Section~\ref{sec:distributed-fusion}) involves inverting or combining $N$ local covariance matrices and summing information vectors. If each node $i$ handles contributions from its neighbors in $\mathcal{N}_i$, the total cost scales with $|\mathcal{N}_i|$. In the worst-case scenario of a fully connected network, each node processes contributions from $N-1$ other nodes. The matrix operations remain $O(n^3)$ per node (for $n$-dimensional state), plus the cost of exchanging local estimates.

Combining the above factors, the runtime at each node per iteration can be approximated as
\begin{equation}
    O\Bigl(\text{NN fwd/bkwd passes} + n^3 + |\mathcal{N}_i|\bigl(n^3 + \text{comm}\bigr)\Bigr),
\end{equation}
where `$\text{comm}$' denotes the cost of communicating local means and covariance matrices (or other fused messages). The exact constants depend on the neural architecture, dimension of the state, and the network connectivity. Nevertheless, since the NDKF operates without collecting raw measurements centrally, it remains more scalable than a fully centralized solution, especially for large $N$ or high-dimensional sensor data.

Per-node compute is dominated by Jacobian evaluations and $O(n^3)$ factorizations; communication scales with node degree $|\mathcal{N}_i|$ (not with $N$ if the graph is sparse). In practice we found that increasing state dimension $n$ calls for \emph{moderately} wider networks to preserve approximation quality (e.g., hidden width scaling sub-quadratically in $n$), while depth increases yield diminishing returns compared to Jacobian-cost growth. Performance degradation with $N$ is typically \emph{sub-linear} when degree is bounded; over-smoothing can be avoided by limiting consensus rounds or using CI. These guidelines help select capacity without excessive complexity.

\section{Convergence and Stability Analysis}
\label{sec:convergence-stability}

This section provides a rigorous analysis of the stability of the NDKF under local linearization and distributed fusion.

\subsection{Assumptions}
The following assumptions are used in the analysis:
\begin{enumerate}[leftmargin=*]
    \renewcommand{\labelenumi}{\roman{enumi}.}
    \item \textit{Local smoothness on a compact operating set:} There exists a compact set $\mathcal{X}\subset \mathbb{R}^n$ that contains the true trajectory and the filter trajectory over the horizon of interest. On $\mathcal{X}$, the mappings $f_{\theta_f}$ and $h_{\theta_{h,i},i}$ are continuously differentiable and locally Lipschitz.
    \item \textit{Bounded Jacobians on $\mathcal{X}$:} There exist constants $M_f,M_{h_i}>0$ such that
    \[
    \sup_{\mathbf{x}\in \mathcal{X}}\|\nabla f_{\theta_f}(\mathbf{x})\| \le M_f,\quad
    \sup_{\mathbf{x}\in \mathcal{X}}\|\nabla h_{\theta_{h,i},i}(\mathbf{x})\| \le M_{h_i}.
    \]
    We do not require global bounds outside $\mathcal{X}$.
    \item \textit{Noise and Initial Error Bounds:} The process noise $\mathbf{w}_k$ and the measurement noise $\mathbf{v}_{k,i}$ are zero-mean and bounded in the mean-square sense. Moreover, there exists a constant $\delta>0$ such that the initial estimation error satisfies
    $\|\mathbf{e}_{0|0,i}\| < \delta$, for all nodes $i$.
    \item \textit{Graph connectivity:} The communication graph is (jointly) connected over the fusion horizon (e.g., $B$-connected).
    \item \textit{Innovation conditioning:} There exists $s_{\min}>0$ such that the local innovation covariance satisfies $\mathbf{S}_{k,i} \succeq s_{\min}\mathbf{I}$ for all $k$ in the operating region (ensured in practice by $\mathbf{R}_i \succ 0$ and measurement excitation).
\end{enumerate}

The Lipschitz/bounded-Jacobian conditions are \emph{enforceable} during training using spectral-norm constraints, gradient penalties, or weight decay; empirical verification can be done by sampling $\|\nabla f_{\theta}\|$ and $\|\nabla h_{\theta,i}\|$ over a validation grid. The bound on $\mathbf{S}_{k,i}$ is a standard detectability/conditioning requirement and can be supported by measurement whitening or adaptive covariance inflation. The inequality $\|\mathbf{I}-\mathbf{K}_{k,i}\mathbf{H}_{\theta,i,k}\|\le \beta_i$ is not \emph{designed} directly (since $\mathbf{K}_{k,i}$ is computed from the Riccati recursion), but in practice can be \emph{promoted} by: (a) maintaining reasonable $\mathbf{R}_i$ (well-conditioned innovations), (b) damping the update $\hat{\mathbf{x}}\leftarrow \hat{\mathbf{x}}+\alpha\,\mathbf{K}\,\mathrm{innov}$ with $\alpha\in(0,1]$, and (c) using CI when correlations are uncertain. Finally, the mean-value linearization remainder can be lumped into an effective process noise, i.e., $\mathbf{Q}\leftarrow \mathbf{Q}+\tilde{\mathbf{Q}}$, when state jumps are large.

\subsection{Local Error Dynamics}
Define the estimation error at node $i$ at time $k$ as
\[
\mathbf{e}_{k|k,i} = \mathbf{x}_k - \hat{\mathbf{x}}_{k|k,i}.
\]
Under local linearization, the prediction step yields an \emph{a priori} error dynamics approximated by
\[
\mathbf{e}_{k+1|k,i} \approx \mathbf{F}_{\theta}(\xi_k)\,\mathbf{e}_{k|k,i} + \mathbf{w}_k,
\]
where $\xi_k$ lies on the line segment between $\mathbf{x}_k$ and $\hat{\mathbf{x}}_{k|k,i}$. Similarly, the measurement update gives
\[
\mathbf{e}_{k+1|k+1,i} \approx \Bigl(\mathbf{I} - \mathbf{K}_{k+1,i}\,\mathbf{H}_{\theta,i,k+1}\Bigr)\,\mathbf{e}_{k+1|k,i} - \mathbf{K}_{k+1,i}\,\mathbf{v}_{k+1,i}.
\]
The distributed fusion step is assumed to combine local estimates (e.g., via an information form aggregation) without degrading these contraction properties.

\subsection{Sufficient Conditions for Stability}

\begin{theorem}[Local mean-square boundedness of NDKF]
\label{thm:local-stability-revised}
Under Assumptions (i)--(v), suppose the true and estimated trajectories remain in
$\mathcal{X}$ and that for each node $i$ there exists a constant $\beta_i \ge 0$ satisfying
\begin{equation}
    \|\mathbf{I} - \mathbf{K}_{k,i}\mathbf{H}_{\theta_{h,i},i,k}\| \leq \beta_i,
    \; \forall\, k,
\;\text{with}\;
    \gamma_i := M_f\,\beta_i < 1.
\end{equation}
Then the local estimation error satisfies
\begin{equation}
\mathbb{E}\!\left[\|\mathbf{e}_{k|k,i}\|^2\right]
\le
C_i\,\gamma_i^{2k}\,\|\mathbf{e}_{0|0,i}\|^2 + \frac{D_i}{1-\gamma_i^2},
\end{equation}
for some constants $C_i,D_i>0$ determined by the noise levels and the linearization remainder on $\mathcal{X}$. Hence NDKF is \emph{locally} exponentially stable in mean square and converges to a bounded noise-dependent neighborhood.
\end{theorem}

\begin{remark}[Interpretation of the assumptions] \mbox{} \\
The bounded-Jacobian condition is a local regularity assumption on the operating region, not a consequence of nor a substitute for global stability. It is practically checkable by
spectral-normalization bounds and by sampling Jacobian norms on a validation grid. When the learned model error increases, its effect can be absorbed conservatively through covariance inflation, which preserves consistency at the cost of larger posterior uncertainty.
Assumptions (iv)–(v) ensure, respectively, network connectivity for distributed fusion and well-conditioned innovations; they are not required for the local contraction bound but are used to preserve performance under distributed operation.
\end{remark}

\begin{proof}
We analyze the error propagation for a single node $i$ and then consider the fusion step. Let $\mathbf{e}_{k|k,i} = \mathbf{x}_k - \hat{\mathbf{x}}_{k|k,i}$ denote the local estimation error at time $k$, and assume that the local linearization is valid for sufficiently small errors.

\noindent\textit{i. Prediction Step.}  
The true state evolves as
\begin{equation}
    \mathbf{x}_{k+1} = f_{\theta}(\mathbf{x}_k) + \mathbf{w}_k,
    \label{eq:true_evolution_revised}
\end{equation}
and the predicted state is given by
\begin{equation}
    \hat{\mathbf{x}}_{k+1|k,i} = f_{\theta}(\hat{\mathbf{x}}_{k|k,i}).
    \label{eq:predicted_state_revised}
\end{equation}
By the mean value theorem, there exists a point $\xi_k$ between $\mathbf{x}_k$ and $\hat{\mathbf{x}}_{k|k,i}$ such that
\begin{equation}
    f_{\theta}(\mathbf{x}_k) - f_{\theta}(\hat{\mathbf{x}}_{k|k,i}) = \mathbf{F}_{\theta}(\xi_k)\,(\mathbf{x}_k - \hat{\mathbf{x}}_{k|k,i}),
    \label{eq:mean_value_revised}
\end{equation}
where $\mathbf{F}_{\theta}(\xi_k)$ is the Jacobian of $f_{\theta}$ at $\xi_k$. By assumption,
\begin{equation}
    \|\mathbf{F}_{\theta}(\xi_k)\| \leq M_f.
\end{equation}
Thus, the \emph{a priori} error satisfies
\begin{equation}
    \|\mathbf{e}_{k+1|k,i}\| \leq M_f\,\|\mathbf{e}_{k|k,i}\| + \|\mathbf{w}_k\|.
    \label{eq:prediction_bound_revised}
\end{equation}

\begin{remark}[Remainder term]
The mean-value form hides a higher-order remainder; under the Lipschitz-Jacobian assumption this remainder is bounded by $O(\|\mathbf{e}_{k|k,i}\|^2)$ and can be conservatively absorbed into an effective process noise, which justifies the inequalities \eqref{eq:prediction_bound_revised}–\eqref{eq:combined_bound_revised}.
\end{remark}

\noindent\textit{ii. Measurement Update.}  
At the measurement update, the filter receives
\begin{equation}
    \mathbf{y}_{k+1,i} = h_{\theta,i}(\mathbf{x}_{k+1}) + \mathbf{v}_{k+1,i},
    \label{eq:measurement_model_proof_revised}
\end{equation}
and updates its state as
\begin{equation}
    \hat{\mathbf{x}}_{k+1|k+1,i} = \hat{\mathbf{x}}_{k+1|k,i} + \mathbf{K}_{k+1,i}\Bigl( \mathbf{y}_{k+1,i} - h_{\theta,i}(\hat{\mathbf{x}}_{k+1|k,i}) \Bigr).
\end{equation}
Define the \emph{a posteriori} error as
\begin{equation}
    \mathbf{e}_{k+1|k+1,i} = \mathbf{x}_{k+1} - \hat{\mathbf{x}}_{k+1|k+1,i}.
\end{equation}
Subtracting the above from the true state \eqref{eq:true_evolution_revised}, linearizing $h_{\theta,i}(\cdot)$ (with Jacobian $\mathbf{H}_{\theta,i,k+1}$ satisfying $\|\mathbf{H}_{\theta,i,k+1}\|\leq M_{h_i}$), and applying assumptions, yields
\begin{multline}
    \|\mathbf{e}_{k+1|k+1,i}\| \leq \|\mathbf{I} - \mathbf{K}_{k+1,i}\mathbf{H}_{\theta,i,k+1}\|\|\mathbf{e}_{k+1|k,i}\| \\+ \|\mathbf{K}_{k+1,i}\|\|\mathbf{v}_{k+1,i}\|.
    \label{eq:measurement_bound_revised}
\end{multline}
By assumption, we have $\|\mathbf{I} - \mathbf{K}_{k+1,i}\mathbf{H}_{\theta,i,k+1}\| \leq \beta_i$, resulting in
\begin{equation}
    \|\mathbf{e}_{k+1|k+1,i}\| \leq \beta_i\left(M_f\|\mathbf{e}_{k|k,i}\| + \|\mathbf{w}_k\|\right) + \|\mathbf{K}_{k+1,i}\|\|\mathbf{v}_{k+1,i}\|.
    \label{eq:combined_bound_revised}
\end{equation}
Defining $\gamma_i = M_f\,\beta_i < 1$, we have the recursion:
\begin{equation}
    \|\mathbf{e}_{k+1|k+1,i}\| \leq \gamma_i\,\|\mathbf{e}_{k|k,i}\| + \beta_i\,\|\mathbf{w}_k\| + \|\mathbf{K}_{k+1,i}\|\,\|\mathbf{v}_{k+1,i}\|.
    \label{eq:recursive_error_revised}
\end{equation}

\noindent\textit{iii. Induction.}  
Iterating \eqref{eq:recursive_error_revised} from $k=0$ to $N$, we have
\begin{multline}
    \|\mathbf{e}_{N|N,i}\| \leq \gamma_i^N\,\|\mathbf{e}_{0|0,i}\| \\+ \sum_{j=0}^{N-1} \gamma_i^{N-1-j}\left(\beta_i\,\|\mathbf{w}_j\| + \|\mathbf{K}_{j+1,i}\|\|\mathbf{v}_{j+1,i}\|\right).
    \label{eq:inductive_bound_revised}
\end{multline}
Since $\gamma_i < 1$, the term $\gamma_i^N\,\|\mathbf{e}_{0|0,i}\|$ decays exponentially, and under the bounded noise assumptions the summation converges. Hence, $\|\mathbf{e}_{N|N,i}\|$ converges exponentially fast to a bounded region determined by the noise covariances.

\noindent\textit{iv. Distributed Fusion Step.}  
After local updates, each node $i$ fuses its estimate with neighbors:
\begin{equation}
    \mathbf{e}^{(\mathrm{fusion})}_{k|k,i} = \sum_{j \in \mathcal{N}_i} w_{ij}\,\mathbf{e}_{k|k,j},
\end{equation}
with weights $w_{ij}\geq 0$ and $\sum_{j\in \mathcal{N}_i}w_{ij}=1$. By convexity,
\begin{equation}
    \|\mathbf{e}^{(\mathrm{fusion})}_{k|k,i}\| \leq \max_{j\in \mathcal{N}_i}\|\mathbf{e}_{k|k,j}\|.
    \label{eq:fusion_bound_revised}
\end{equation}
Thus, the fusion step preserves the contraction properties established above.

Collectively, there exist constants $C, C' > 0$ such that
\begin{equation}
    \mathbb{E}\Bigl[\|\mathbf{e}_{k|k,i}\|^2\Bigr] \leq C\,\gamma_i^{2k}\,\|\mathbf{e}_{0|0,i}\|^2 + \frac{C'}{1-\gamma_i^2}.
\end{equation}
This establishes the exponential mean-square stability of the NDKF under the stated conditions.
\end{proof}

\subsection{Practical Proxies and Optimality}
Instead of verifying $\|\mathbf{I}-\mathbf{K}\mathbf{H}\|\le \beta$ directly, practitioners can monitor: (i) innovation conditioning ($\lambda_{\min}(\mathbf{S}_{k,i})\ge s_{\min}$), (ii) bounded covariance ($\operatorname{tr}(\mathbf{P}_{k|k})\le p_{\max}$), and (iii) Jacobian norms ($\|\mathbf{F}_{\theta}(\cdot)\|\le M_f$, $\|\mathbf{H}_{\theta,i}(\cdot)\|\le M_{h_i}$) induced by training constraints. These proxies imply a contraction margin in the linearization region and are straightforward to track online.

With exact models, Gaussian noise, and valid local linearization, NDKF reduces locally to a distributed information filter/EKF and is locally MMSE on the linearization region. With learned and therefore imperfect models, NDKF is generally \emph{suboptimal}; in that regime, the more relevant goal is consistency, which can be promoted through covariance inflation or covariance-intersection fusion.

\section{Results and Discussion}
\label{sec:results}

In this section, we present empirical evaluations of the proposed NDKF on a 2D system monitored by four distributed sensor nodes, and compare its performance to an Extended Kalman Filter (EKF) baseline.

\subsection{Experimental Setup}
\label{sec:exp-setup}

\noindent\textit{System Simulation:}  
We consider a 2D state vector 
\(
\mathbf{x}_k = [p_x,\, p_y]^{T},
\)
which evolves according to
\begin{equation}
    \mathbf{x}_{k+1} = \mathbf{x}_k + 
    \begin{bmatrix}
        0.05 \cos\bigl(k/10\bigr), 
        0.05 \sin\bigl(k/10\bigr)
    \end{bmatrix}^\top
    + \boldsymbol{\omega}_k,
\end{equation}
where $\boldsymbol{\omega}_k \sim \mathcal{N}\bigl(\mathbf{0},\,\mathrm{diag}(0.001,0.001)\bigr)$. Here, $\mathcal{N}(\cdot,\cdot)$ denotes the multivariate normal (Gaussian) distribution with the specified mean and covariance matrix. Offline training data are generated over 400 time steps, and an independent set of 100 time steps is used for testing and filter evaluation.

\noindent\textit{Measurement Models:}  
Each node obtains a single 1D measurement of the 2D state, implying partial observability at the individual node level:
\begin{itemize}
    \item[] {Node 1:} $z_{k,1} = \sin(2p_x) + 0.5\,p_y + \nu_{k,1}$,
    \item[] {Node 2:} $z_{k,2} = \cos(2p_y) - 0.4\,p_x + \nu_{k,2}$,
    \item[] {Node 3:} $z_{k,3} = \sin(2p_x) + \cos(2p_y) + \nu_{k,3}$,
    \item[] {Node 4:} $z_{k,4} = \sin(2p_x) - \cos(2p_y) + \nu_{k,4}$.
\end{itemize}
Here, $\nu_{k,i} \sim \mathcal{N}(0,0.01)$. In the NDKF implementation, each node’s measurement function $h_{\theta,i}(\cdot)$ is learned from data using a dedicated neural network.
The EKF baseline employs mis‑specified analytical measurement functions for Node 1 and 2 (i.e., ignoring the additional scaling and bias terms to assess performance under model mismatch), while using the correct functions for Node 3 and 4.

\noindent\textit{Neural Network Configuration and Training Process:}  
To train the neural networks used in the NDKF, we adopt an offline learning strategy. First, we collect datasets consisting of state transition pairs $\{(\mathbf{x}_k^{(j)}, \mathbf{x}_{k+1}^{(j)})\}_{j=1}^{M}$ for the dynamics network and measurement pairs $\{(\mathbf{x}_k^{(j)}, \mathbf{y}_{k,i}^{(j)})\}_{j=1}^{M_i}$ for each sensor node's measurement network. These datasets are split into training and validation sets. The dynamics network $f_{\theta}(\cdot)$ is implemented as a deep fully connected network with three hidden layers (128 neurons each), with batch normalization and dropout (0.2) to prevent overfitting. It is trained for 3000 epochs using the Adam optimizer with an initial learning rate of 0.001 and a learning rate scheduler. Similarly, each measurement network $h_{\theta,i}(\cdot)$ is implemented with two hidden layers (32 neurons each) and trained for 1000 epochs using analogous settings. The training objective minimizes the mean squared error loss functions in \eqref{eq:dyn-loss} and \eqref{eq:meas-loss}. Once convergence is achieved, the learned parameters are fixed and subsequently used in the online filtering process.

\noindent\textit{Filter Initialization:}  
All nodes are initialized with 
\[
\hat{\mathbf{x}}_{0|0,i} = \begin{bmatrix}0, \, 0\end{bmatrix}^{T} \quad \text{and} \quad \mathbf{P}_{0|0,i} = 0.5\,\mathbf{I}_2.
\]
The process noise covariance is set to $\mathbf{Q}=\mathrm{diag}(0.001,0.001)$ and the measurement noise covariance to $\mathbf{R}_i=0.01$ for all nodes.

\subsection{Experimental Evaluation}
We evaluate filtering performance using the \emph{root mean squared error} (RMSE):
\vspace{-5pt}
\begin{equation}
    \mathrm{RMSE}(k) \;=\; \Big(\frac{1}{N_t}\sum_{t=1}^{N_t} \|\mathbf{x}_t - \hat{\mathbf{x}}_{t|t}\|^2\Big)^{\frac{1}{2}},
\end{equation}
computed at each time step $k$ over $N_t$ Monte Carlo runs. Additionally, we analyze the measurement innovation residuals to assess the quality of the update step.
Table~\ref{tab:rmse-comparison} summarizes the RMSE for estimating $p_x$ and $p_y$ averaged over 40 runs.
These results indicate that the proposed NDKF achieves approximately a 70\% reduction in $p_x$ error and a 41\% reduction in $p_y$ error compared to the EKF baseline.

The improvement in Table~\ref{tab:rmse-comparison} is primarily attributable to reduced \emph{model mismatch}, not to a different Kalman algebra. In the benchmark, the EKF baseline uses mis-specified observation functions at Nodes~1 and~2, whereas NDKF learns these mappings directly from data. As a result, the NDKF innovations are better centered and the local
posteriors fused across nodes are less biased. We therefore interpret the improvement as evidence that learned models can be advantageous when analytical process or measurement equations are incomplete or inaccurate.

\begin{table}[!b]
\caption{RMSE Comparison for $p_x$ and $p_y$ (40 Monte Carlo Runs)}
\label{tab:rmse-comparison}
\small
\centering
\setlength{\tabcolsep}{5pt}
\begin{tabular}{lcc}
\hline
\textbf{Method} & \textbf{RMSE in $p_x$} & \textbf{RMSE in $p_y$} \\
\hline
NDKF (proposed) & 0.1209 & 0.2642 \\
Distributed EKF (baseline) & 0.3964 & 0.4463 \\
\hline
\end{tabular}
\normalsize
\end{table}

\begin{figure}[!tb]
    \centering
    \includegraphics[trim={20pt 10.5cm 22pt 65pt},clip,width=\columnwidth]{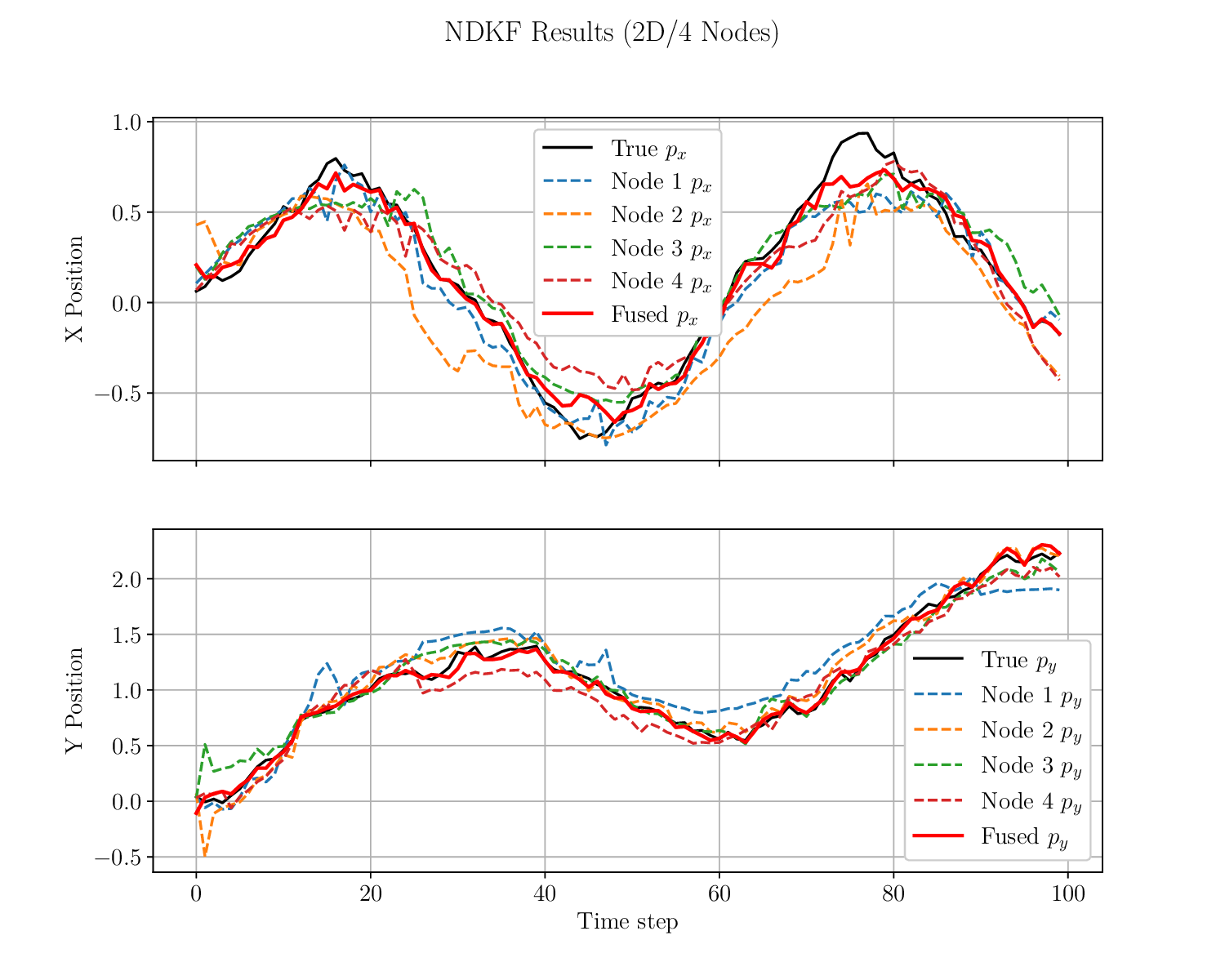}
    \includegraphics[trim={20pt 28pt 22pt 10.5cm},clip,width=\columnwidth]{plots/trajectory_plot.eps}
    \caption{True state, individual node estimates, and the fused NDKF estimate.}
    \label{fig:trajectory}
    \vspace{-7pt}
\end{figure}

\begin{figure}[!tb]
    \centering
    \includegraphics[trim={20pt 0pt 20pt 32pt},clip,width=\columnwidth]{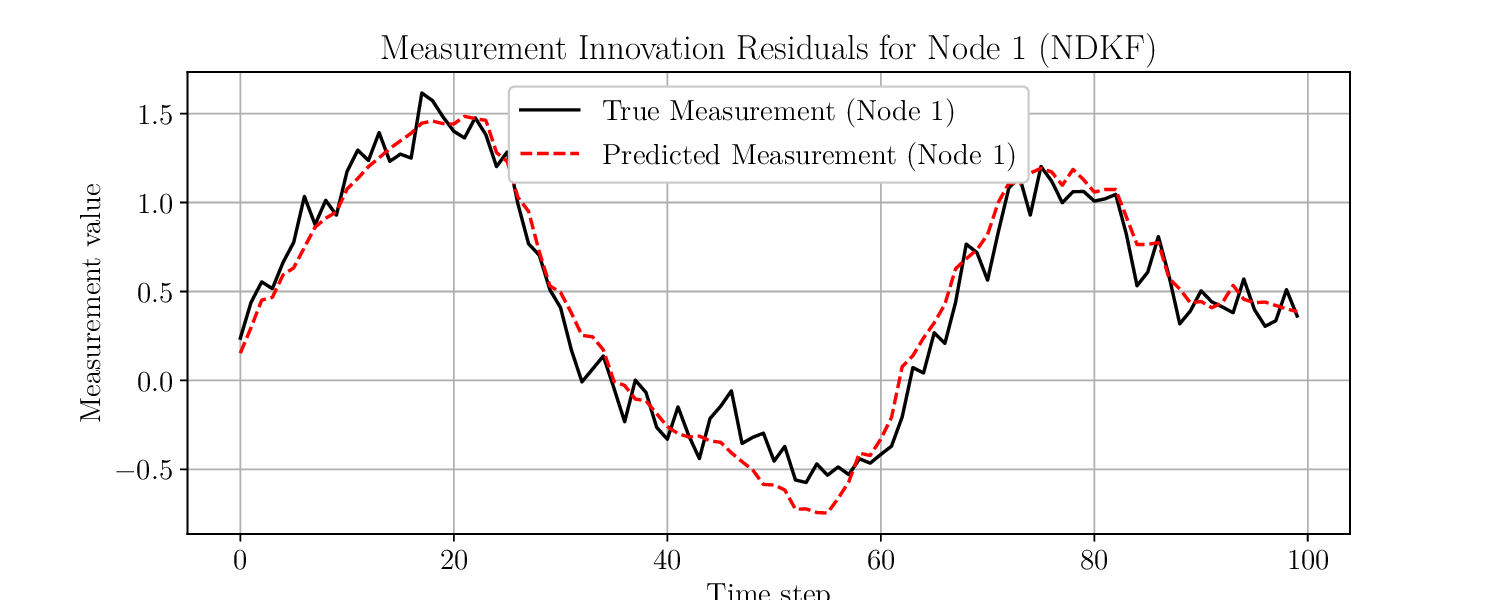}
    \caption{Measurement innovation residuals for Node 1.}
    \label{fig:innovation}
\end{figure}

Fig.~\ref{fig:trajectory} shows a representative 2D trajectory comparing the true state, the local node estimates, and the fused state estimate obtained using the NDKF. The NDKF’s fused estimate closely tracks the true state, while the EKF baseline deviates considerably, as indicated by the numerical RMSE in Table~\ref{tab:rmse-comparison}, due to its mis‑specified measurement models. Fig.~\ref{fig:innovation} illustrates the measurement innovation residuals for Node 1. The low and stable residuals obtained with the NDKF, along with the numerical results reported in Table~\ref{tab:rmse-comparison}, confirm its superior measurement update performance compared to the EKF baseline.\looseness=-1


The experimental results show that in scenarios with significant nonlinearities and model mismatch, the proposed NDKF outperforms the distributed EKF baseline. By learning both the dynamics residual and the true measurement functions, the NDKF achieves much lower RMSE values (0.1209 for $p_x$ and 0.2642 for $p_y$) compared to the EKF (0.3964 and 0.4463, respectively), even though the EKF uses the correct measurement models for Nodes 3 and 4. While the EKF may perform well under near‑linear conditions, its reliance on fixed, mis‑specified analytical models leads to significant errors when facing true nonlinear behavior.

An implementation of NDKF, including the example provided in this section, is available in the GitHub repository accompanying this paper at: {\small\url{github.com/sfarzan/NDKF}}

\section{Conclusion and Future Directions}\label{sec:conclusion}

We introduced the Neural-Enhanced Distributed Kalman Filter (NDKF), a data-driven framework that integrates learned neural network models within a consensus-based, distributed state estimation scheme. By replacing traditional process and measurement equations with learned functions $f_{\theta}(\cdot)$ and $h_{\theta,i}(\cdot)$, NDKF is able to accommodate complex or poorly characterized dynamics while remaining robust in the presence of partial and heterogeneous observations.
Our theoretical analysis identified key factors influencing filter convergence, including Lipschitz continuity of the neural approximations, proper noise covariance tuning, and adequate network connectivity for effective distributed fusion, along with a detailed computational complexity analysis. Experimental evaluations on a 2D system with multiple sensor nodes demonstrated the NDKF’s capability to reduce estimation error compared to a baseline distributed EKF, particularly under challenging nonlinear motion patterns.

Future work will study larger-scale systems, different communication graphs, distribution shifts, outlier-contaminated measurements; and investigate mode-conditioned or hybrid learned models for non-smooth dynamics.\looseness=-1

\bibliographystyle{IEEEtran}
\bibliography{bibtex}

\newpage
\begin{appendix}
This appendix elaborates on practical NDKF training strategies, including residual learning with nominal models and self-supervised learning from innovations, and discusses how model accuracy, online adaptation, and non-smooth operating regimes affect deployment and interpretation of the stability results. It also presents a lightweight innovation-gating mechanism for improving robustness to outliers and faulty sensors.

\subsection{Residual Learning with Nominal Dynamics and Measurements}
When approximate physics-based models $f_0$ and $h_{0,i}$ are available, we learn only
corrections:
\begin{align}
f_{\theta_f}(\mathbf{x}) &= f_0(\mathbf{x}) + \Delta f_{\theta_f}(\mathbf{x}), \\
h_{\theta_{h,i},i}(\mathbf{x}) &= h_{0,i}(\mathbf{x}) + \Delta h_{\theta_{h,i},i}(\mathbf{x}).
\end{align}
Using proxy states $\tilde{\mathbf{x}}_k$ obtained from a smoother, calibration system, or
digital twin, we optimize
\begin{multline}
\mathcal{L}_{\mathrm{res\mbox{-}dyn}}
=
\sum_{k}
\Bigl\|
\tilde{\mathbf{x}}_{k+1}-f_0(\tilde{\mathbf{x}}_k)-\Delta f_{\theta_f}(\tilde{\mathbf{x}}_k)
\Bigr\|^2
\\[-5pt] +\lambda_f\sum_k \|\Delta f_{\theta_f}(\tilde{\mathbf{x}}_k)\|^2,
\end{multline}
\vspace{-10pt}
\begin{multline}
\mathcal{L}_{\mathrm{res\mbox{-}meas},i}
=
\sum_{k}
\Bigl\|
\mathbf{y}_{k,i}-h_{0,i}(\tilde{\mathbf{x}}_k)-\Delta h_{\theta_{h,i},i}(\tilde{\mathbf{x}}_k)
\Bigr\|^2
\\[-5pt] +\lambda_{h,i}\sum_k \|\Delta h_{\theta_{h,i},i}(\tilde{\mathbf{x}}_k)\|^2.
\end{multline}
The residual penalties keep the learned correction small when the nominal model is already
informative, which helps preserve filter consistency.

\paragraph*{Self-supervised learning from innovations}
When state labels are unavailable, we treat the state sequence as latent and train the models
by minimizing the negative predictive log-likelihood induced by the filter. Let
\[
\boldsymbol{\nu}_{k,i}
=
\mathbf{y}_{k,i}-h_{\theta_{h,i},i}(\hat{\mathbf{x}}_{k|k-1}), \]
\[ \mathbf{S}_{k,i}
=
\mathbf{H}_{\theta_{h,i},i,k}\mathbf{P}_{k|k-1}\mathbf{H}_{\theta_{h,i},i,k}^{\top}
+\mathbf{R}_i.
\]
We then minimize
\begin{equation}
\mathcal{L}_{\mathrm{ss}}(\theta_f,\{\theta_{h,i}\})
=
\sum_{k=1}^{T}\sum_{i=1}^{N}
\left(
\boldsymbol{\nu}_{k,i}^{\top}\mathbf{S}_{k,i}^{-1}\boldsymbol{\nu}_{k,i}
+
\log\det \mathbf{S}_{k,i}
\right)
\label{eq:selfsup-loss}
\end{equation}
optionally together with Jacobian regularization terms such as
\[
\lambda_J\sum_k \|\nabla f_{\theta_f}(\hat{\mathbf{x}}_{k|k-1})\|_F^2
\]
to promote smoothness. In practice, we alternate short filter rollouts with gradient updates
and stop training when the validation innovation statistics cease to improve.

\paragraph*{Effect of model accuracy and online adaptation}
Let $\delta_f(\mathbf{x}) := f_{\star}(\mathbf{x}) - f_{\theta_f}(\mathbf{x})$ denote the
dynamics-model mismatch. In the local error recursion, $\delta_f(\cdot)$ enters as an
additional disturbance term and can therefore be absorbed into an inflated effective process
covariance as long as the mismatch and Jacobian error remain bounded on the operating region.
Consequently, better dynamics models enlarge the practical region where the contraction proxy
$\gamma_i<1$ is observed, whereas poorer models reduce this margin and may require stronger
covariance inflation or shorter adaptation intervals. If distribution shift is predominantly
sensor-specific, updating the local $h_{\theta_{h,i},i}$ models is typically sufficient. If
the shared process model drifts, the updated $\theta_f$ need not be exchanged at every
filtering step; instead, it can be synchronized only when a new model checkpoint is deployed,
since the online communication layer transmits posterior summaries rather than network
parameters.

\paragraph*{Non-smooth or switching dynamics}
The analysis in Section~\ref{sec:convergence-stability} is local and applies on compact
operating regions where the effective dynamics are well approximated by a smooth model. For
piecewise-smooth or mode-switching systems, a single Lipschitz-constrained network may blur
sharp transitions. In such cases, a practical extension is to use a bank of mode-conditioned
networks or a mixture-of-experts model with smooth gating, and to run the NDKF within each
mode or on the active expert. We therefore do not claim that a single smooth network is
universally adequate for arbitrary hybrid dynamics; rather, the present framework targets
regimes that are locally smooth over the estimation horizon.

\paragraph*{Model-class dependence}
Although the presentation focuses on neural networks, the local stability argument relies
primarily on differentiability on the operating region, bounded Jacobians, and well-conditioned
innovations, not on the specific choice of function approximator. The same blueprint can
therefore be adapted to alternative models such as Gaussian-process mean predictors or RBF
networks, provided their mean maps admit bounded derivatives on the region of interest and the
induced linearization can be evaluated reliably.

\subsection{Outlier Handling and Faulty Sensors}
Covariance intersection is used in NDKF as a consistency-preserving fusion rule when inter-node
cross-correlations are unknown; it is \emph{not} by itself an outlier-rejection mechanism. To
improve robustness to spurious measurements, each node can gate its local update using the
normalized innovation squared (NIS)
\begin{align}
\eta_{k+1,i}
&=
\boldsymbol{\nu}_{k+1,i}^{\top}\mathbf{S}_{k+1,i}^{-1}\boldsymbol{\nu}_{k+1,i}, \\
\boldsymbol{\nu}_{k+1,i}
&=
\mathbf{y}_{k+1,i}-h_{\theta_{h,i},i}\bigl(\hat{\mathbf{x}}_{k+1|k}\bigr),
\end{align}
and compare $\eta_{k+1,i}$ with a $\chi^2$ threshold $\tau_{\alpha,m_i}$. If
$\eta_{k+1,i}>\tau_{\alpha,m_i}$, node $i$ either skips the update or inflates the local noise
model according to
\begin{equation}
\mathbf{R}_i \leftarrow \kappa_i \mathbf{R}_i, \qquad \kappa_i > 1.
\end{equation}

This introduces one additional step in Algorithm 1 presented in section \ref{subsec:alg}:
\begin{itemize}
\item[5)] \textit{Optional gating:} compute
$\eta_{k+1,i}=\boldsymbol{\nu}_{k+1,i}^{\top}\mathbf{S}_{k+1,i}^{-1}\boldsymbol{\nu}_{k+1,i}$.
If $\eta_{k+1,i}>\tau_{\alpha,m_i}$, skip the update or set
$\mathbf{R}_i \leftarrow \kappa_i \mathbf{R}_i$ before recomputing $\mathbf{K}_{k+1,i}$.
\end{itemize}

This addition is lightweight, preserves the distributed structure, and is compatible with
covariance-intersection fusion.

\end{appendix}

\end{document}